# Modeling Behavior Change for Multi-model At-Risk Students Early Prediction (extended version)

Jiabei Cheng
Department of Computing
The Hong Kong Polytechnic University
Hong Kong, China
jiabcheng@polyu.edu.hk

Zhen-Qun Yang
Department of Computing
The Hong Kong Polytechnic University
Hong Kong, China
zq-cs.yang@polyu.edu.hk

Jiannong Cao
Department of Computing
The Hong Kong Polytechnic University
Hong Kong, China
jiannong.cao@polyu.edu.hk

Yu Yang
Department of Computing
The Hong Kong Polytechnic University
Hong Kong, China
cs-yu.yang@polyu.edu.hk

Kai Cheung Franky Poon
HKRSS Tai Po Secondary School
Hong Kong, China
kcpoon@hkrsstpss.edu.hk

Daniel Lai
Department of Computing
The Hong Kong Polytechnic University
Hong Kong, China
daniel.sc.lai@hkjc.org.hk

*Abstract*— In the educational domain, identifying students at risk of dropping out is essential for allowing educators to intervene effectively, improving both academic outcomes and overall student well-being. Data in educational settings often originate from diverse sources, such as assignments, grades, and attendance records. However, most existing research relies on online learning data and just extracting the quantitative features. While quantification eases processing, it also leads to a significant loss of original information. Moreover, current models primarily identify students with consistently poor performance through simple and discrete behavioural patterns, failing to capture the complex continuity and non-linear changes in student behaviour. We have developed an innovative prediction model, Multimodal- ChangePoint Detection (MCPD), utilizing the textual teacher remark data and numerical grade data from middle schools. Our model achieves a highly integrated and intelligent analysis by using independent encoders to process two data types, fusing the encoded feature. The model further refines its analysis by leveraging a changepoint detection module to pinpoint crucial behavioral changes, which are integrated as dynamic weights through a simple attention mechanism. Experimental validations indicate that our model achieves an accuracy range of 70- 75%, with an average outperforming baseline algorithms by approximately 5-10%. Additionally, our algorithm demonstrates a certain degree of transferability, maintaining high accuracy when adjusted and retrained with different definitions of at-risk, proving its broad applicability.

## I. Introduction

Elementary and secondary school stages are pivotal for student development. Personal and environmental factors during these periods significantly influence students, leading to potential learning obstacles and negative behaviour habits. These challenges threaten their overall growth and increase the risk of dropping out, which affects the quality of education and the school's social responsibilities. Thus, effective analysis and timely intervention are essential to identify and support at-risk students. By doing so, schools can enhance educational standards, lower dropout rates, uphold their reputation, and protect student rights.

Analyzing and predicting at-risk students typically requires reliance on multi-source data, including academic performance, behavioural characteristics, and psychological states, to ensure the accuracy of the predictions. Despite the raw data having different types, the existing methods for predicting at-risk students can broadly be divided into two categories based on the approach to feature extraction: predictions based on numerical features [1]–[6] and those based on multi-modal features [7]–[9]. Predictions based on numerical features involve converting data from various modalities into numerical features (e.g., frequency of activities, difficulty of tasks) through manual or automated methods, such as neural networks. This approach is simple and direct. On the other hand, predictions based on multi-modal features employ different types of neural networks to extract features from various data types (such as text, images, etc.), which are then fused [10], [11]. While much of the research on predicting at-risk students focuses on using numerical features due to their ease of processing, it is crucial to acknowledge that relying solely on data from a single modality, particularly numerical data, often yields limited information.

The majority of studies utilizing multi-model features focus on online data [7], [8], [12]–[14]. This phenomenon is mainly due to the proliferation of online courses, which provide data such as student login times and course viewing frequencies. This data is not only abundant and clean but also covers a wide range of types, facilitating the extraction of more accurate and comprehensive features. However, primary and secondary education mostly uses offline teaching with online registration to track student performance. This approach often leads to chaotic raw data, making the extracted features less reflective of students' real situations [15], [16]. Obtaining higher-quality features in an offline educational setting might require the deployment of the Internet of Things (IoT) technologies such as cameras and sensors. However, this approach can significantly limit the amount and coverage of the data collected, directly affecting the quality of the features.

It is noteworthy that while many studies recognize the importance of student behavioral patterns [16] for predicting at-risk students, most treat student behavior in isolation, overlooking the continuity of behavior [18]–[21], [28]. Our analysis finds that student behavior typically exhibits linear and smooth changes, where any sudden and dramatic variation could indicate the student is undergoing a difficult period, potentially leading to a collapse in their belief system and, consequently, a decision to drop out. Therefore, monitoring changes in academic performance and related behavioral habits is critical for educational analysis systems. This effort aims to enhance the prediction of at-risk students and ensure timely identification and correction of negative changes. However, research in this area is relatively scarce.

This paper aims to predict the at-risk students in the next academic year using existing offline multi-source, multi-modal behavioral and academic data from middle schools. We propose a new prediction method, Multimodal-ChangePoint Detection (MCPD), which is constructed using deep learning and combines text processing and time series analysis. Features are extracted from academic data and teacher comments through the MCPD model and then fused to predict at-risk students. An attention mechanism based on changepoint detection is integrated to monitor abrupt changes in student behavior, thereby enhancing prediction accuracy. The innovations of this paper include:

- Develop a multi-modal model that effectively combines numerical and textual data from students' academic performance and teachers' comments to preserve original data and efficiently fuse diverse inputs;
- Incorporating an attention module based on changepoint detection to determine student behavior change;
- Validated the effectiveness and flexibility of the model.

## II. RELATED WORKS

Previously, online systems combined semantic search for analysis and pattern identification for student data, however, semantic search also has its limitations [22], [23]. Predicting risks such as student dropout, withdrawal, and psychological issues requires a comprehensive examination of multi-source and multi-modal information, including academic performance, learning habits, and behavioral characteristics. Researchers are also concerned about the process and utilization of existing data to obtain compelling features.

M. Young [24] employs surveys to gather information on academics, family, health, etc., converting this information into numerical data through manual scoring. The information obtained remains quite limited due to limitations in survey design and information conversion methods. A framework named CASTLE for detecting college students' psychological health issues was introduced by [8], which develops a specialized algorithm called MOON to integrate multi-modal data like social life, academic performance, and physical appearance for gathering students' social relationship data. Video of students' classroom performance is collected using cameras, and times when students face forward and write are semi-automatically extracted as a measure of attentiveness during class to comprehensively predict college students' future academic achievements [25]. For elementary students' learning and behavioral characteristics, the Wearables and Internet of Things for Education (WIoTED) system is designed by [9], collecting multi-modal information such as class dates, duration of participation in learning and activities, organizers of activities, and the number of interactions during activities, and converting it into numerical data. Using statistical analysis tools, this conversion significantly improves the accuracy of student learning outcome predictions.

Predicting at-risk students is recognized as a sequence labelling or time series forecasting issue, where applying LSTM models to capture temporal information can effectively enhance at-risk prediction accuracy, as emphasized by [26]. The article's CLSA model includes a mutual attention mechanism based on features at different time points, which integrates temporal information into features to improve at-risk prediction performance. It is posited by [27] that students' learning sequence information essentially constitutes time-series data, with varying intervals between events, hence proposing a time-controlled Long Short-Term Memory neural network (ELSTM) prediction model capable of modelling early learning behaviours over different time intervals. The full utilization of the temporality of student information, employing prediction models established through logistic regression with regularization terms and an Input-Output Hidden Markov Model (IOHMM) to predict students at risk of dropping out early, is aimed at by [20]. Integrating existing theories with data analysis, L. Wang et al. [19] highlight the strong correlation between students' phases of behavioral changes and their eventual dropout outcomes, identifying it as one of the most critical indicators for predicting students at risk of dropping out. Therefore, based on the insights from this article, we incorporate sudden changes in student behavior as a critical component of our prediction model, enhancing our ability to forecast students' dropout risk accurately.

In our model, we utilized changepoint detection [29], a method grounded in statistical analysis, to evaluate whether behaviors at specific time points constitute sudden changes. Change-point detection is a crucial technique in signal processing for identifying structural shifts in time-series data, known as 'change points' [30]. These points often indicate potential state transitions or anomalous events and have wide applications across various fields, including network traffic monitoring and biomedical signal processing. Therefore, in analyzing educational data, we employ change-point detection to identify significant changes in student behavior.

## III. DATA MINING

In this section, we specifically extracted data related to types of behavioral descriptions, including reasons for student absences, descriptions of participation in activities, reasons for punishments, and reasons for awards. We compared the significant differences between at-risk students and non-at-risk students. Moreover, we underscored the importance of multi-modal features by comparing the information before and after data quantification.

### A. Differences in Numerical Data

Analyzing behavioural data numerically—by counting be- behaviour occurrences—indicates that at-risk students experience a greater number of difficulties compared to their non-at-risk peers. From Fig. 1, it's evident that at-risk students have higher and more varied counts of punishments, absences and relatively fewer rewards, highlighting the disparities in their experiences. The variation in these counts suggests significant differences, yet it does not conclusively show whether at-risk students have distinct patterns of activity involvement.

### B. Differences in original Text Data

Thus, we have listed descriptions that **only appear in the behavior of at-risk students** in the activity and punishment tables. In the first word cloud shown in Fig. 2 (left), We can observe that certain vocabulary indicates numerous serious behaviours among at-risk students. such as "smoking", "touching body", "attacking", and "fighting". Additionally, words emphasizing frequency, such as 'multiple times' and "repeatedly uncorrected", suggest these students repeatedly break the rules and are difficult to amend through regular advice. These findings align with our understanding of at-risk students.

In the second word cloud shown in Fig. 2 (right), we observe terms like 'art', 'sports', and 'pet grooming' that seem unrelated to the main academic content of the school. This suggests that the interests of these at-risk students might not lie in their studies. Instead, extracurricular activities could consume a significant amount of their time on non-academic pursuits, leading to greater challenges in their academic work.

**Direct exposure to textual data reveals that semantics can better describe student statuses, an aspect not represented in numerical data, reaffirming the necessity of a multimodal approach.**

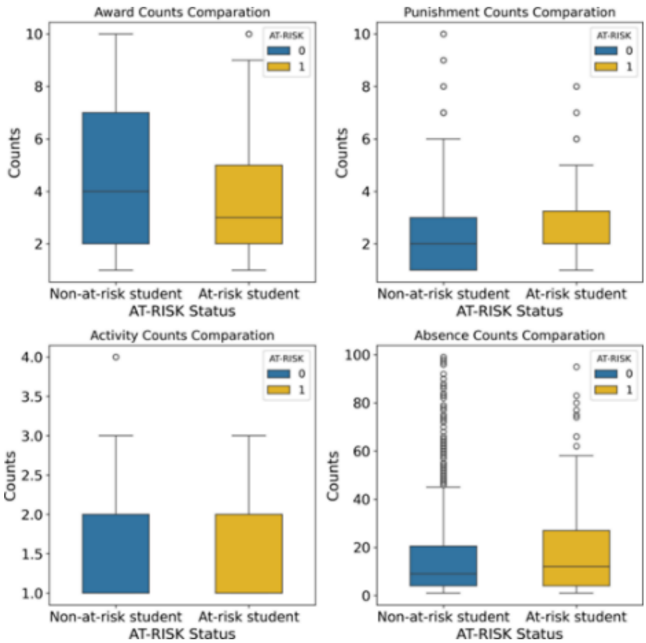

Fig. 1. Differences in Numerical Data.

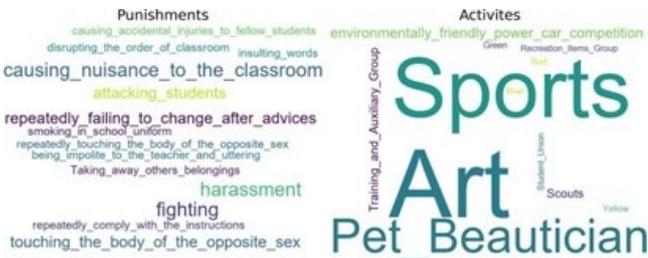

Fig. 2. Differences in Punishment and Activity Data.

## IV. METHOD

### A. Dataset

Our study data originate from a secondary school located in the urban area of Hong Kong, encompassing detailed records of 1,248 students within the year from 2019 to 2022. Within this dataset, 128 students have been classified as at-risk-those who dropped out in the following year (labelled as 1, otherwise labelled as 0). Our dataset is divided into two main categories: numerical and textual data. The data consists of numerical and textual parts. Numerical data records students' grades in Chinese, Mathematics, and English across six exams per year. Textual data details students' behaviours at school, including activity participation, reasons for absences, rewards and punishment. The year is divided into six periods aligning with the exams, and behaviors are summarized for each period. For example, 'During this period, the student was absent X times, with reasons including..., received Y rewards, for reasons such as..., faced Z punishments, due to..., and participated in N activities, involving...'.

### B. Model Design

In this study, we developed a new method called the Multimodal-ChangePoint Detection (MCPD) Model, which is shown in Fig. 3, to predict students' dropout risk. This model assesses risk based on students' behavioral patterns and academic performance over the year.

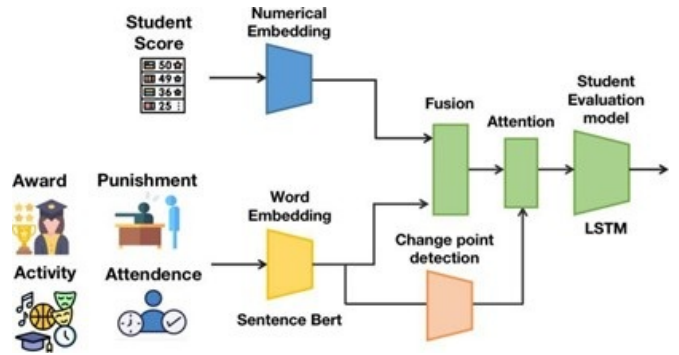

Fig. 3. Structure of the Multimodal-ChangePoint Detection (MCPD) Model.

The model integrates Sentence-BERT [31], LSTM [32], and changepoint detection technologies to effectively extract features from textual behavior descriptions and academic records. For numerical data, a simple two-layer feedforward neural network was employed to extract key academic performance features. For textual data, the Sentence-BERT model was utilized to transform complex behavioral descriptions into deep semantic features.

Subsequently, we fused these two types of features by directly concatenating them within each period to form a comprehensive behavioral feature vector. This vector not only expresses the student's academic achievements but also reflects deeper psychological states regarding their learning status. Moreover, our study first introduces the Bayesian changepoint detection algorithm in predicting at-risk students. Bayesian change point detection is a statistical method that infers when structural changes occur in a data sequence by calculating changes in the probability distributions before and after potential change points in the observed data. We applied this algorithm to the behavioral vectors of students for each period, identifying potential abrupt changes in behavior. These probabilities were then used as weights, similar to an attention mechanism, multiplied by the behavioral features to highlight significant changes in behavioral patterns.

The weighted comprehensive behavioral feature vectors were then fed into an LSTM model. With its ability to capture long-term dependencies, the LSTM analyzed hidden patterns and dynamics within the time series data, thereby enhancing the accuracy of dropout risk prediction. The final predictive output is generated through a sigmoid activation function, indicating the probability of student dropout.

## V. EXPERIMENT

Our Multimodal-ChangePoint Detection (MCPD) Model, aimed at predicting at-risk middle school students, shows significant efficacy by integrating diverse data types, including textual and numerical data, with changepoint detection for behavioral changes. This highlights the model's robust capability in handling complex educational data and adaptability across various at-risk student definitions. The MCPD model offers a comprehensive approach to identifying students facing academic and social challenges, paving the way for targeted interventions to improve student outcomes and well-being.

### A. Model Effectiveness

To evaluate the performance of the model proposed in this study, we adopted baseline methods for comparative analysis of prediction accuracy. Like most existing studies, the baseline models convert textual data into numerical features and then merge with other numerical data to form a comprehensive feature set for prediction. Specifically, these comprehensive features are applied to a variety of traditional machine learning models, including Long Short-Term Memory Networks (LSTM), Support Vector Machines (SVM), Logistic Regression, Adaboost, and Random Forests, to predict at-risk students.

For performance evaluation, we used accuracy (ACC), recall, and F1-score as the key metrics. These evaluation metrics provide a comprehensive perspective on model performance, considering the accuracy of predictions and the ability to identify students at risk of dropping out. The experimental results are detailed in Table I. Comparative analysis clearly shows that the model proposed by the MCPD model significantly outperforms baseline models across evaluation metrics. This advantage is attributed to the efficiency of our model in feature extraction: by combining Sentence-BERT, changepoint detection technology, and LSTM, our model can more deeply mine and utilize the rich features contained within textual data and time-series data.

TABLE I. COMPARISON OF MODEL PERFORMANCE

| Model | Accuracy | Recall | F1 Score |
|---|---|---|---|
| **MCPD** | **0.75** | **0.73** | **0.74** |
| LSTM | 0.71 | 0.68 | 0.67 |
| SVM | 0.67 | 0.65 | 0.66 |
| Adaboost | 0.66 | 0.65 | 0.63 |
| Random Forest | 0.63 | 0.609 | 0.64 |
| Logistic Regression | 0.53 | 0.51 | 0.52 |

### B. Model Flexibility

In the methodology section of our study, we considered the broad applicability and transferability of our model across different educational settings. Through collaboration with front-line educational workers, we discovered that teachers need to identify at-risk students beyond just those at risk of dropping out. Thus, we expanded our classification system to include academic crises, severe behavioural crises, and time management crises, detailed in Table II. This expansion meets the needs of secondary school teachers and tests the broad applicability of our education method. The specific approach is as follows: we updated the target labels to reflect these newly defined risk categories, thereby enhancing the predictive capabilities of our model. Additionally, we adjusted key hyperparameters to better match the nuanced distinctions of the new risk definitions and retrained the model using a revised dataset with updated annotations. Experimental results demonstrate that our model now achieves an accuracy rate exceeding 70% in predicting these newly defined at-risk student groups, showcasing its enhanced ability to address a broader range of student issues.

TABLE II. EXPLANATION OF CRISIS TYPES

| Name | Explanation | Criterion | Accuracy |
|---|---|---|---|
| Academic Crisis | A significant decline in grades reflects that a student may be facing challenges in their studies. | Student's grades fall into the bottom 15% of the year more than 3 times. | 73.6% |
| Severe Behavioral Crisis | Severe behavioural issues in students often indicate the need for psychological health and behavioural interventions. | The student has been judged for major and minor misconducts. | 71.2% |
| Time Management Crisis | Excessive involvement in extracurricular activities may hinder a student's academic performance. | Student join more than 2 activities and ranks in the bottom 15% of the entire grade. | 65.3% |

## VI. CONCLUSION

Implementing the Multimodal-ChangePoint Detection (MCPD) Model has significantly improved the accuracy of identifying at-risk middle school students, achieving over 75% accuracy. This model adapts well across various definitions of at-risk students, demonstrating broad applicability. The integration of changepoint detection to note abrupt behavioral changes has been crucial in enhancing prediction accuracy.

Despite its promising results, the MCPD model faces limitations such as performance variability across different educational contexts and reliance on accurate, comprehensive behavioural data, which may not always be available. Future research could investigate the long-term outcomes of interventions based on MCPD predictions, and developing a more interactive model for real-time predictions by educators could also greatly increase the model's utility in educational environments.

ACKNOWLEDGMENT

This work is supported by Hong Kong Jockey Club Charities Trust (Project S/N Ref.: 2021-0369), and the Research Institute for Artificial Intelligence of Things, The Hong Kong Polytechnic University.